\documentclass[aps,prb,twocolumn]{revtex4}
\usepackage{graphicx}% Include figure files

\begin{document}

\title{Trapping Abelian anyons in fractional quantum Hall droplets}

\author{Zi-Xiang Hu}
\author{Xin Wan}
\affiliation{Zhejiang Institute of Modern Physics, Zhejiang University,
Hangzhou, 310027, P.~R. China}

\author{Peter Schmitteckert}
\affiliation{Institut f\"ur Nanotechnologie,
   Forschungszentrum Karlsruhe, D-76021 Karlsruhe,
   Germany}

\date{\today}

\begin{abstract}
  We study the trapping of Abelian anyons (quasiholes and
  quasiparticles) by a local potential (e.g., induced by an AFM tip)
  in a microscopic model of fractional quantum Hall liquids with
  long-range Coulomb interaction and edge confining potential. We
  find, in particular, at Laughlin filling fraction $\nu = 1/3$, both
  quasihole and quasiparticle states can emerge as the ground state of
  the system in the presence of the trapping potential.  As expected,
  we find the presence of an Abelian quasihole has no effect on the edge
  spectrum of the quantum liquid, unlike in the non-Abelian case
  [Phys. Rev. Lett. {\bf 97}, 256804 (2006)].  Although quasiholes and
  quasiparticles can emerge generically in the system, their stability
  depends on the strength of the confining potential, the strength and
  the range of the trapping potential. We discuss the relevance of the
  calculation to the high-accuracy generation and control of
  individual anyons in potential experiments, in particular, in the
  context of topological quantum computing.

\end{abstract}

\maketitle

\section{Introduction}

Shortly after the discovery of the fractional quantum Hall
effect~\cite{PhysRevLett.48.1559},
Laughlin~\cite{PhysRevLett.50.1395} realized that electrons in such
a system form an incompressible quantum liquid with excitations of
fractional charge.~\cite{PhysRevLett.52.1583,PhysRevLett.53.722}
These exotic particle excitations~\cite{leinaas77,wilczek82} are
dubbed (Abelian) anyons. To interchange two anyons, one obtains a
phase factor $e^{i
  \theta}$ for the wave function, where $\theta$ is neither an
integral multiple of $2 \pi$ as required by bosons, nor an odd
multiple of $\pi$ as required by fermions. The presence of these
particles with fractional statistics is an indication of topological
phases.~\cite{wenbook} So far, experiments have confirmed the
fractional
charge,~\cite{Science.267.1010,Nature.389.162,PhysRevLett.79.2526} but
the direct observation of the fractional statistics remain
questionable.~\cite{camino:246802,camino:075342,goldman:045334,kim:216404,jain:136802,rosenow07}
Recent experiments~\cite{camino:246802,camino:075342} demonstrated the
so-called superperiods in the conductance oscillations in a fractional
quantum Hall quasiparticle interferometer, which appear to be
consistent with fractional
statistics.~\cite{goldman:045334,kim:216404} However, some theoretical
works~\cite{jain:136802,rosenow07} raised subtleties in the
interpretations.

A second family of anyons is believed to exist in the
fractional quantum Hall state at $\nu = 5/2$. The even-denominator
state is belived to be a $p$-wave paired state, known as the
Moore-Read state or the Pfaffian state, which supports half-flux
quantum vortex excitations.~\cite{moore91} 
Such particle excitations carry $e/4$
charge each and, when interchanged, not just add a phase factor to
the wave function, but evolve unitarily in its degenerate (or
quasi-degenerate for finite systems) ground state manifold. They are dubbed
non-Abelian anyons, which are also speculated to exist at $\nu =
12/5$. The existence of the non-Abelian anyons, although not confirmed by
experiments yet, is of vital importance to topological quantum
computing.~\cite{Kitaev,freedman02a,freedman02b,Preskill}

In theory, the wave functions of quasihole excitations can be
written explicitly in analytic functions for both the Laughlin case
and the Moore-Read case. They are also exact eigenstates of some
special Hamiltonians with short-range two-body and three-body
interactions, respectively. Exact diagonalization of finite systems
has fruitfully revealed some of these quasihole/quasiparticle
states.~\cite{PhysRevLett.54.237} In systems with Coulomb
interaction, such ground state descriptions appear to be sufficient
even for electrons on a Corbino disk geometry in Abelian
cases,~\cite{tsiper:076802} as well as for electrons on a disk
geometry in the non-Abelian case at $\nu = 5/2$.~\cite{wan06} In the
latter case, the change of the edge spectrum in the presence of an
odd number of non-Abelian anyons at the origin implies the
non-Abelian statistics of such excitations. In addition, up to four
non-Abelian quasiholes have been induced and oriented tetrahedrally
on a sphere, which results in two nearly degenerate states with very
similar charge density profile (presumably a topologically protected
qubit).~\cite{haldaneunpublished}

To achieve fault-tolerant quantum computing in the topological
fashion, one needs to be able to create individual, paired, or a
small cluster of anyons. One of the simplest experimental approaches
is probably to use a biased AFM (atomic force microscopy) tip to
create and trap anyons. One may then easily move the anyons
localized at the tip to realize braiding to fulfill computation.
However, the feasibility of creating anyons at an AFM tip has not
yet been systematically studied even on the numerical level.  In a
earlier work by one of the authors and collaborators,~\cite{wan06}
it is demonstrated that a short-range repulsive local potential (as
produced by a sharp AFM tip) can induce both $+e/4$ and $+e/2$
quasiholes, depending on the potential strength, in a $\nu = 5/2$
system. However, a mixture of long-range Coulomb interaction and
short-range three-body interaction is used, and it is not clear
whether negatively charged quasiparticles can be created in a
similar fashion.

In this work, we study the excitation and trapping of both quasiholes
and quasiparticles with a local potential in a microscopic model of
fractional quantum Hall droplets with both long-range Coulomb
interaction and realistic edge confining potential. We focus on the
Laughlin primary filling fraction $\nu = 1/3$, although the approach
can be applied to other filling fractions, including the intriguing
$\nu = 5/2$ case,~\cite{wan06,wan07} to obtain similar results. We
find that both positively charged quasiholes and negatively charged
quasiparticles can be excited generically by a finite-range tip
potential with appropriate sign and strength. We confirm that edge
spectrum of the system is not affected by the presence of a single
quasihole, characteristic of its Abelian nature. Our results suggest
it is possible to trap individual anyons, as needed in 
topological quantum computer proposals.  We also discuss the
stability of anyons when the strength of the confining potential
varies.

The rest of the paper is organized as follows. In Sec. II, we
consider the short-range hard-core potential, which generates the
Laughlin state and the single-quasihole state as exact
zero-energy ground states.  We consider long-range Coulomb
interaction in Sec. III, where we apply tip potentials of
$\delta$-function, Gaussian, and exponential forms. We summarize our
results and discuss the relevance to experiments in the context of
topological quantum computing in Sec. IV.

\section{Hard core interaction}

\begin{figure}
\includegraphics[width=4.5cm]{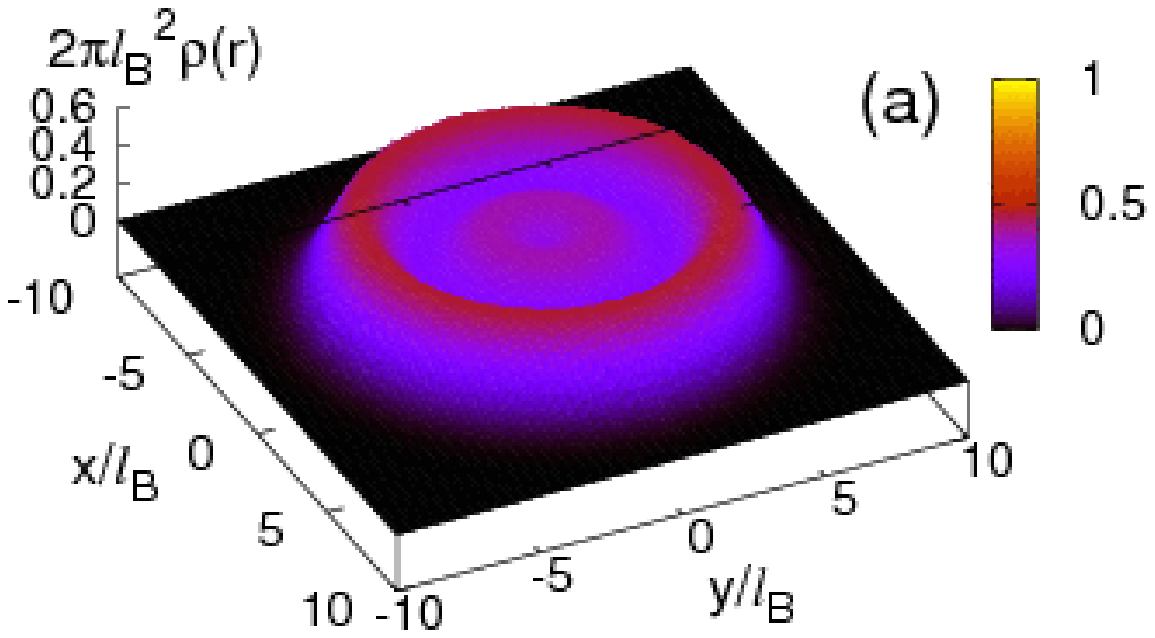}
\includegraphics[width=3.5cm]{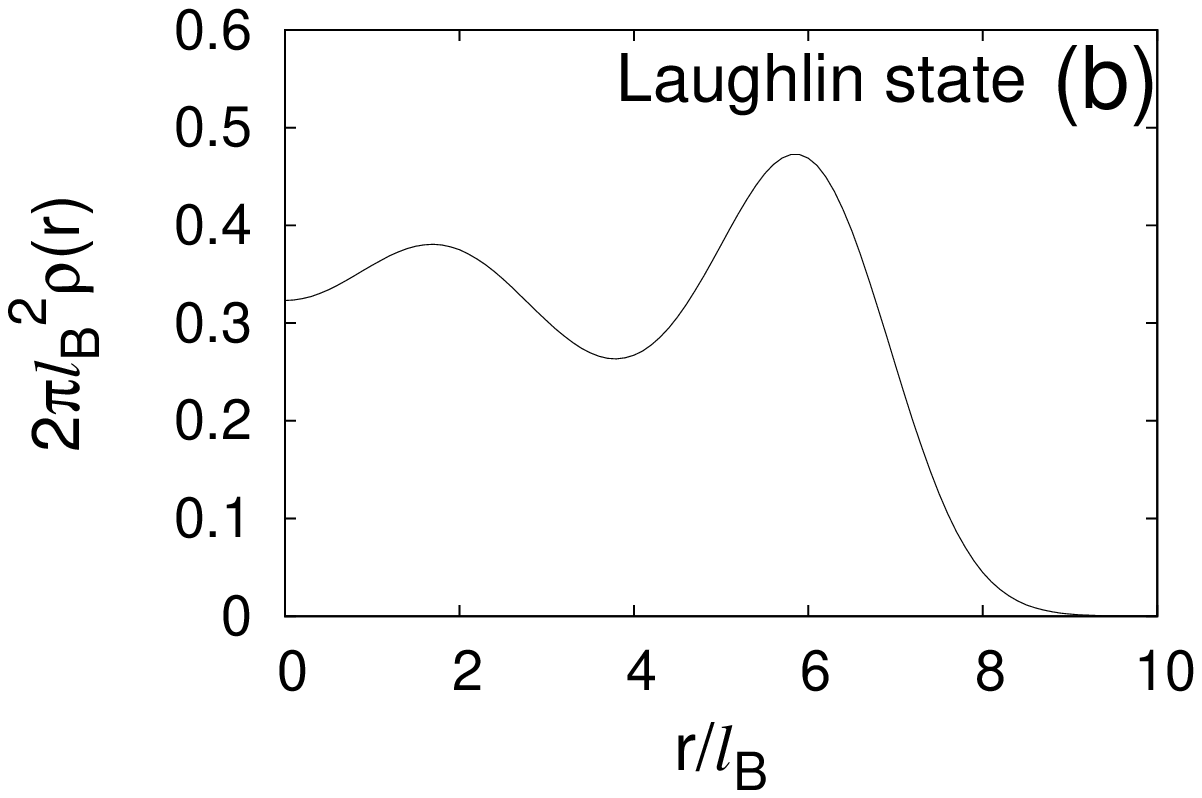} \\
\includegraphics[width=4.5cm]{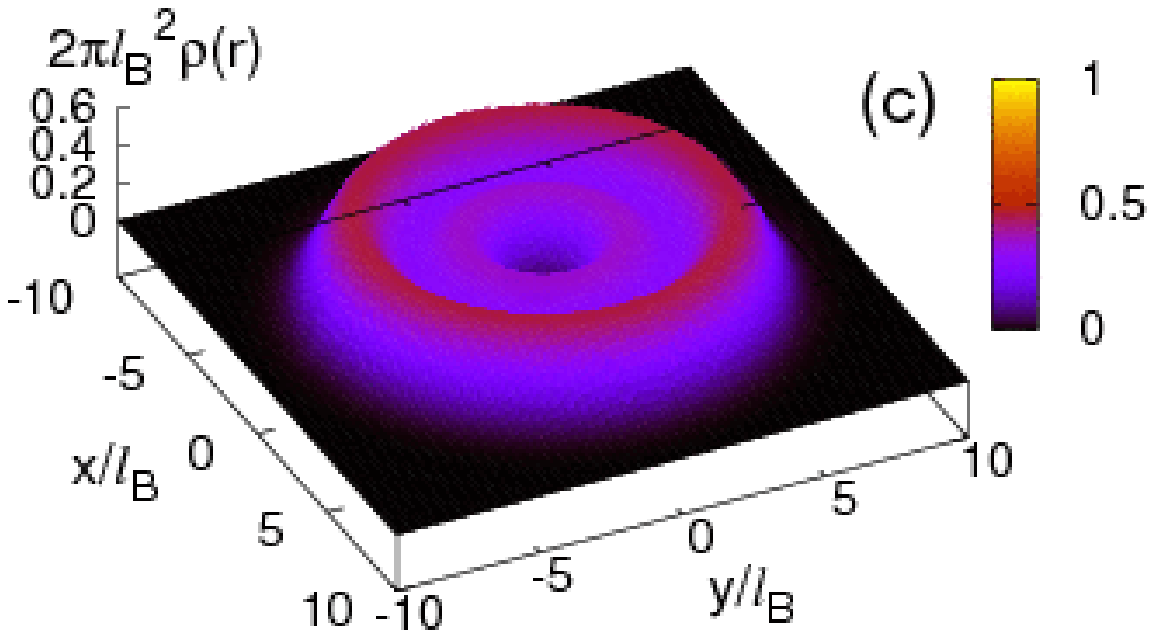}
\includegraphics[width=3.5cm]{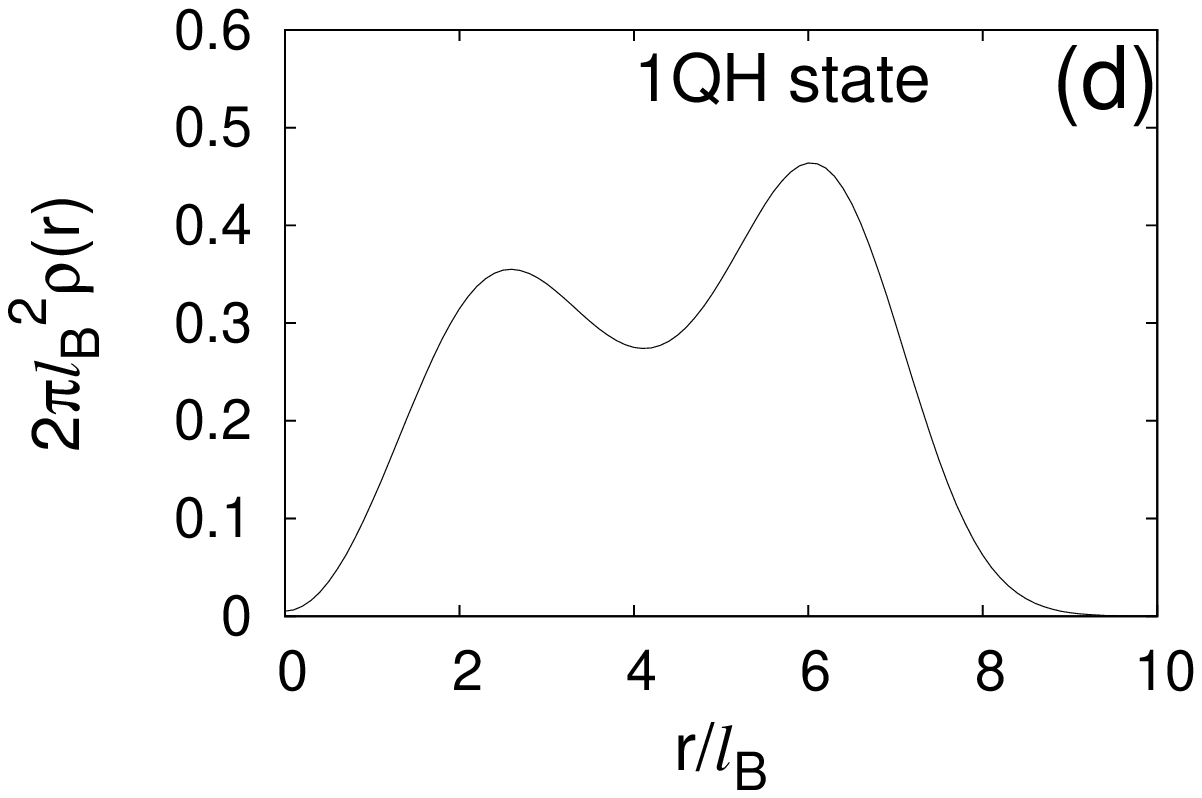} \\
\includegraphics[width=4.5cm]{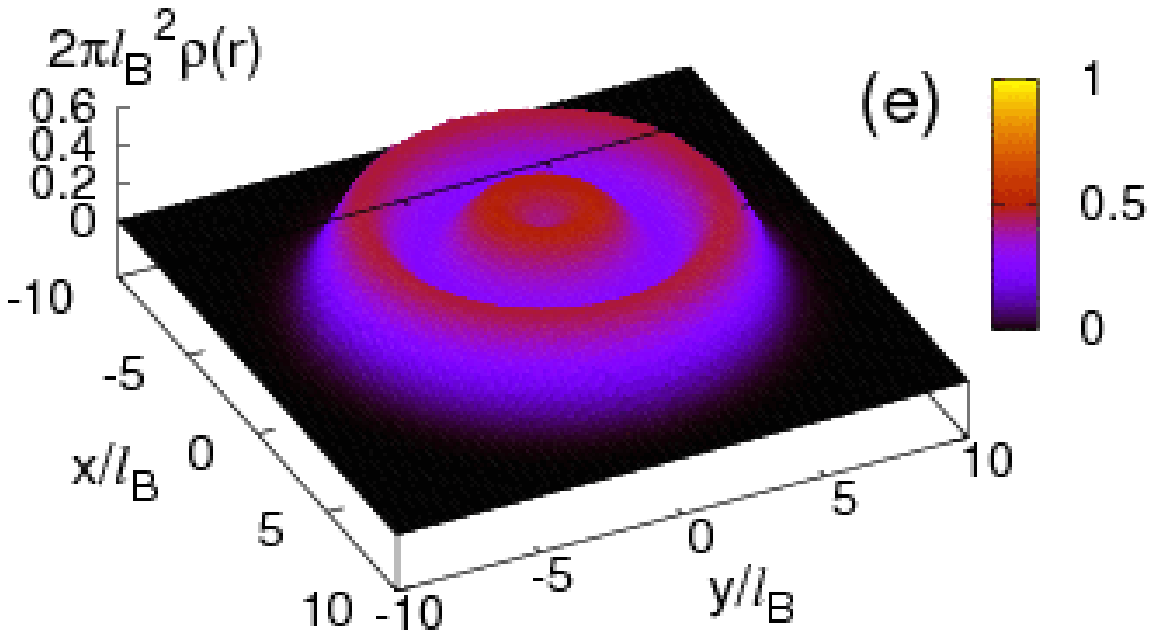}
\includegraphics[width=3.5cm]{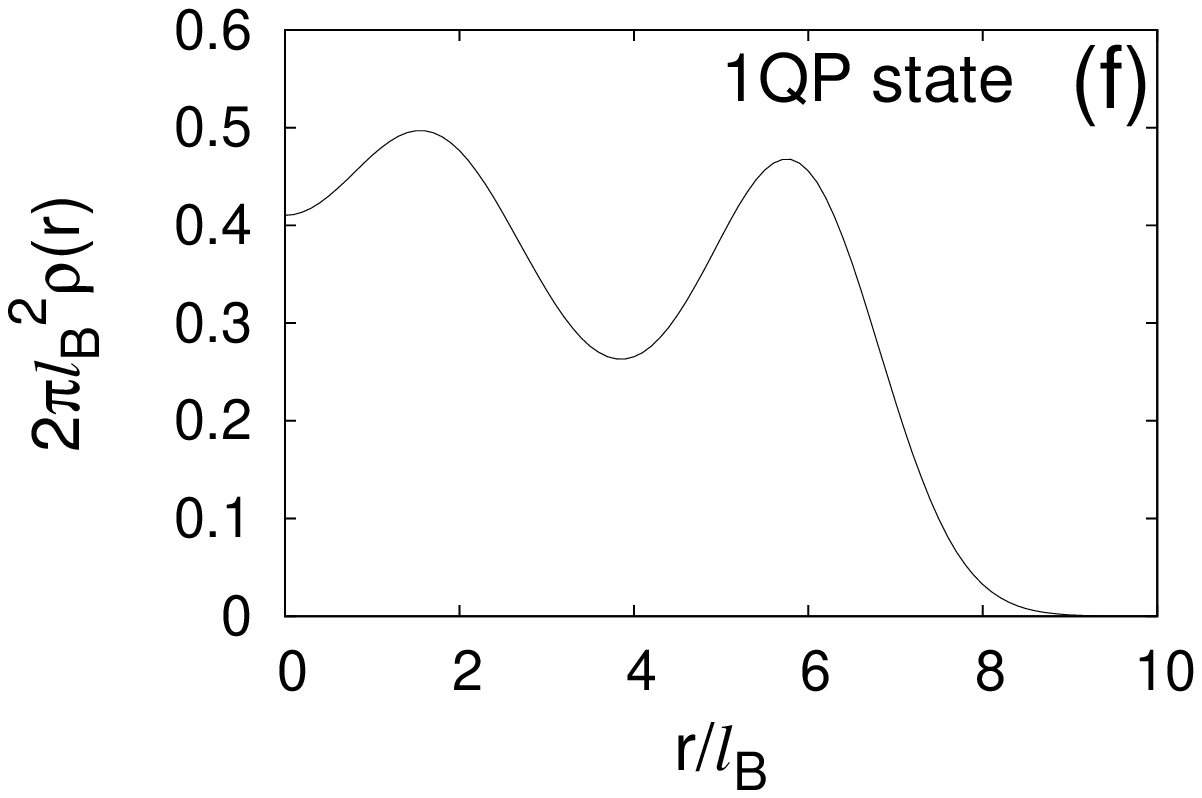}
\caption{\label{v1profile} (color online). The density profiles of
  the 2D electrons with hard-core interaction.
  The Fig. (a) and (b) are the density of Laughlin state.
  Fig. (c) and (d) describe the quasihole state while we add an
  external potential $W_0c_0^{\dagger}c_0$ at m=0 orbit($W_0=0.1$).
  Fig. (e) and (f) are for the quasiparticle candidate state, which looks in the same shape
  as in ref.~\cite{PhysRevLett.54.237}.}
\end{figure}

In this section, we study the two-dimensional electron system on a
disk at filling fraction $\nu = 1/3$ with short-range hard-core
interaction between electrons. In Haldane's pseudopotential language,
$V_m = \delta_{1,m}$. The Laughlin state~\cite{PhysRevLett.50.1395} at
the primary filling factor $\nu = 1/3$
\begin{equation}\label{laughlin}
\Psi _{1/3} (z_1 \cdots z_N ) = \prod\limits_{i > j}^N {(z_i  - z_j
)^3 } \exp \left \{ - \frac{1}{4}\sum\limits_{i = 1}^N {|z_i |^2 }
\right \}
\end{equation}
is the exact ground state with zero energy in the subspace with
total angular momentum $M_{tot} = M_L = 3N(N-1)/2$ for $N$ electrons
in at least $N_{orb}=3N-2$ orbitals. In fact, it is the zero-energy
ground state with the smallest allowed angular momentum; other
zero-energy states (for larger $N_{orb}$) are known as edge states.
We plot the density profile of the Laughlin state for 10 electrons
in 28 orbitals using palette-mapped 3D plot in Fig.~\ref{v1profile}(a) and, for comparison, along the radial direction in
Fig.~\ref{v1profile}(b).

As Laughlin pointed out, the state with a single quasihole at $\xi$
can be written as
\begin{equation}\label{eqn:quasihole}
\Psi _{1/3}^{qh} (\xi; z_1 \cdots z_N ) = \prod\limits_{i = 1}^N
{(z_i  - \xi )} \Psi _{1/3} (z_1 \cdots z_n).
\end{equation}
In the disk geometry, $\xi$ can be placed at the origin to perserve
rotational symmetry. Obviously, this is a zero-energy ground state
in the $M_{1qh} = 3N(N-1)/2 + N$ momentum subspace for $N_{orb} >
3N-2$. In general, there can be additional zero-energy states in the
same momentum subspace, with the wave function being the Laughlin
state multiplied by a symmetric polynomial of order $N$. Such
degeneracy can be lifted either by limiting $N_{orb}=3N-1$, or by
the addition of an impurity potential $H_W = Wc_0^{\dagger}c_0$ at
the $m = 0$ orbital. In Fig.~\ref{v1profile}(c) and Fig.~\ref{v1profile}(d), we plot the density profile of the quasihole wave function.
A density deficiency around the origin is clearly visible,
indicating the presence of a quasihole roughly the size of one
magnetic length $l_B$.

On the other hand, the quasiparticle state of the corresponding
Laughlin state is of some ambiguity. There is no zero-energy state
obtained in the exact diagonalization at $M_{tot} = M_{1qp} =
3N(N-1)/2 - N$, bacause the excitation gap of the Laughlin liquid is
finite. Here, it is hard to compare it with the variational
quasiparticle wave function
\begin{eqnarray}\label{eqn:quasiparticle}
&&{\Psi}_{1/3}^{qp} (\xi, z_1 \cdots z_N ) \nonumber \\
&=&\prod\limits_{i = 1}^N
\left [ e^{-|z_i |^2 /4} \left (2{\partial \over \partial z_i}  - \xi^* \right )
e^{|z_i |^2 /4} \right ] \Psi _{1/3}(z_1 \cdots z_n)
\end{eqnarray}
proposed by Lauglin, which is not known as the exact solution of any
simple Hamiltonian. We assume that, like the quasihole state, the
quasiparticle state is the ground state of the Hamiltonian of
interested at the appropriate angular momentum $M_{1qh}$. Here we
plot the density profile as such a candidate for a quasiparticle
state in Fig.~\ref{v1profile}(e) and Fig.~\ref{v1profile}(f).

\begin{figure}
\includegraphics[width=8cm]{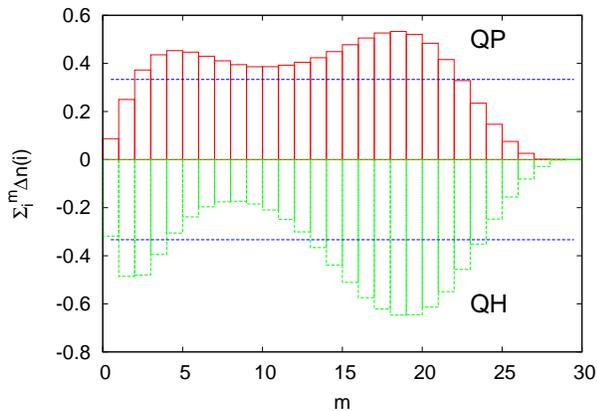}
\caption{\label{v1profilenum} (color online).  The accumulated
difference of the electron occupation number $\sum_{i=0}^m \Delta
n(i)$ between quasihole/quasiparticle state and the Laughlin state
which oscillates around 1/3 (-1/3) for quasihole (quasipaticle)
state, indicating the emergence  of a charge +e/3 (-e/3) quasihole
(quasiparticle)}
\end{figure}

We plot the accumulated difference of the electron occupation numbers
$\sum_{i=0}^m \Delta n(i) = \sum_{i=0}^m \left [n^{qh,qp}(i) - n^L (i)
  \right ]$ between quasihole/quasiparticle state (with electron
occupation number $n^{qh}$ or $n^{qp}$) and the Laughlin state (with
electron occupation number $n^{L}$) in Fig.~\ref{v1profilenum}. The
dotted line in this figure is the average value, {\it i.e.}, 1/3 (or
$-1/3$) for the quasihole (or quasiparticle) state. This confirms that
there are $\pm e/3$ charged excitations in a Laughlin liquid of $\nu =
1/3$. In the case of a hard-core potential, the size of a $-e/3$
charged quasiparticle ($\sim 3 l_B$) is larger than that of a $+e/3$
charged quasihole ($\sim l_B$).

A similar density-profile plot of the Laughlin quasihole and
quasiparticle states in spherical geometry has been reported in an
earlier numerical work.~\cite{PhysRevLett.54.237} Studies on the
Laughlin quasiparticle excitations in the disk geometry with long-range
Coulomb interaction (in the presence of neutralizing background charge
to be realistic) have been absent.

\section{Coulomb interaction}
In this section we study the excitations in a Laughlin liquid of
$N$ electrons with Coulomb interaction, confined by uniform
neutralizing background charge (on a disk of radius $R$) at a
distance $d$ above the electron layer. We use the disk geometry with
the symmetric gauge $\vec{A}=(-\frac{B y}{2}, \frac{B x}{2})$, the
single-particle wave function $\phi_m$ in the lowest Laudau level
is:
\begin{equation}
\phi_m(z) = (2 \pi 2^m m!) ^{-1/2} z^m e^{-|z|^2/4}
\end{equation}
where $z = x + iy$ is the complex coordinate in the electron layer.
Projected to the lowest Laudau level, the Hamiltonian in the second
quantization language reads
\begin{equation}
  H_C = \frac{{\rm{1}}}{{\rm{2}}}\sum\limits_{{\rm{mnl}}}
  {V_{mn}^l c_{m + l}^ +  c_n^ +  c_{n + l} c_m }  +
  \sum\limits_m {U_m c_m^ +  c_m }
\end{equation}
where $c_m^+$ ($c_m$) creates (annihilates) an electron at the
$m$-th orbital. $V_{mn}^l$ are Coulumb matrix elements
\begin{equation}
  V_{mn}^l  = \int {d^2 r_1 \int {d^2 r_2 \phi _{m + l}^*
(\vec{r}_1 )\phi
      _{n}^* (\vec{r}_2 )\frac{{e^2 }}{{\varepsilon r_{12} }}
\phi _{n+l} (\vec{r}_2
      )\phi _m (\vec{r}_1 )} },
\end{equation}
and $U_m$ the background confining potential
\begin{equation}
  U_m = {N e^2 \over \pi R^2 \varepsilon } \int d^2 r \int_{\rho  < R}
d^2 \rho \frac{|\phi _m (\vec{r})|^2}{\sqrt{|\vec{r} - \vec{\rho}|^2 + d^2}}.
\end{equation}

In order to study the quasiparticle and quasihole excitations, we
include an external local potential $H_W$, created by an AFM tip,
for example. So the complete Hamiltonian is
\begin{equation}
H = H_C + H_W.
\end{equation}
In the following, we will consider three different forms of
$H_W$: (i) a short-range potential at the origin of the disk $H_W
= W_0c_0^{\dagger}c_0$; (ii) a Gaussian potential $H_W = W_g
\sum_m \exp (-m^2 / 2 s^2)c_m^{\dagger}c_m$; and (iii) an
exponential potential $H_W = W_e \sum_m \exp (-m /
\xi)c_m^{\dagger}c_m$.

\subsection{Short-range potential at origin}

A short-range potential can be produced by a very sharp AFM tip. By
sharp we mean the range of the tip potential on the 2DEG is smaller
than one magnetic length, the size of a single-particle wave
function in the lowest Landau level. In this case, we can model the
potential by $H_W = W_0 c_0^{\dagger} c_0$, located at the origin in
our disk geometry. A previous study~\cite{wan06} has applied the
short-range potential to create a single $+e/4$ quasihole and two
$+e/4$ quasiholes (or a $+e/2$ quasihole) in a model of the
fractional quantum Hall liquid at $\nu = 5/2$ with Coulomb
interaction and an edge confining potential.

To begin with, we apply the same short-range potential $H_W$ to the
electron liquid at $\nu = 1/3$. We present the results of a system of
$N = 8$ electrons in 26 orbitals (large enough so that edge
excitations have low enough energies). The background charge is still
confined to a disk of $R = \sqrt{2N/\nu} = \sqrt{48}$, corresponding
to the lowest 24 orbitals, at a distance $d = 0.5 l_B$ above the
electron layer.  We expect the ground state of Laughlin nature has a
total angular momentum of $M_L = 3N(N-1)/2 = 84$, which is found to be
right for zero and small $W_0$. When we increase $W_0$ above $0.26 \pm
0.01$, the total angular momentum of the global ground state jumps
from 84 to 92, indicating the excitation of a $+e/3$ quasihole. The
density profile of the quasihole state is similar to that found for
the hard-core potential (Fig.~\ref{v1profile}c and d), in which the
electron density approaches zero at the origin.

\begin{figure}
\includegraphics[width = 8cm]{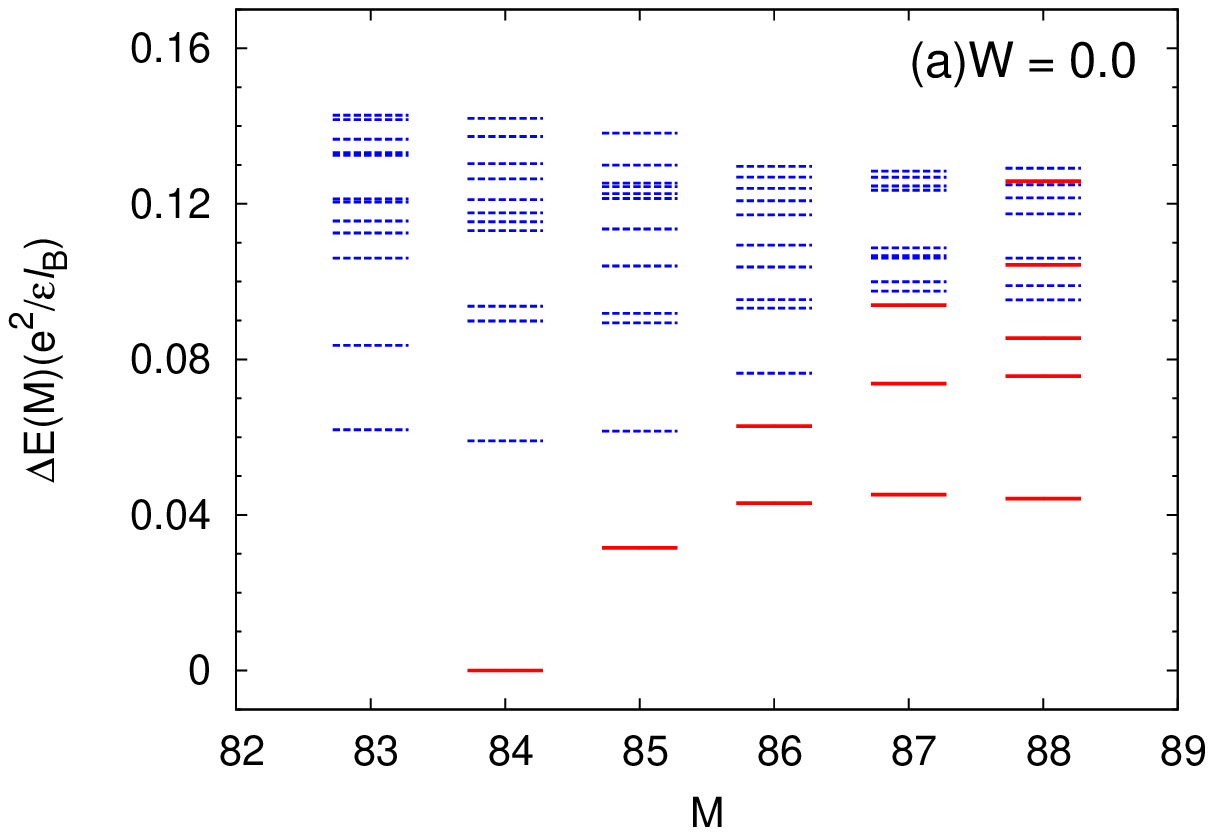}
\includegraphics[width = 8cm]{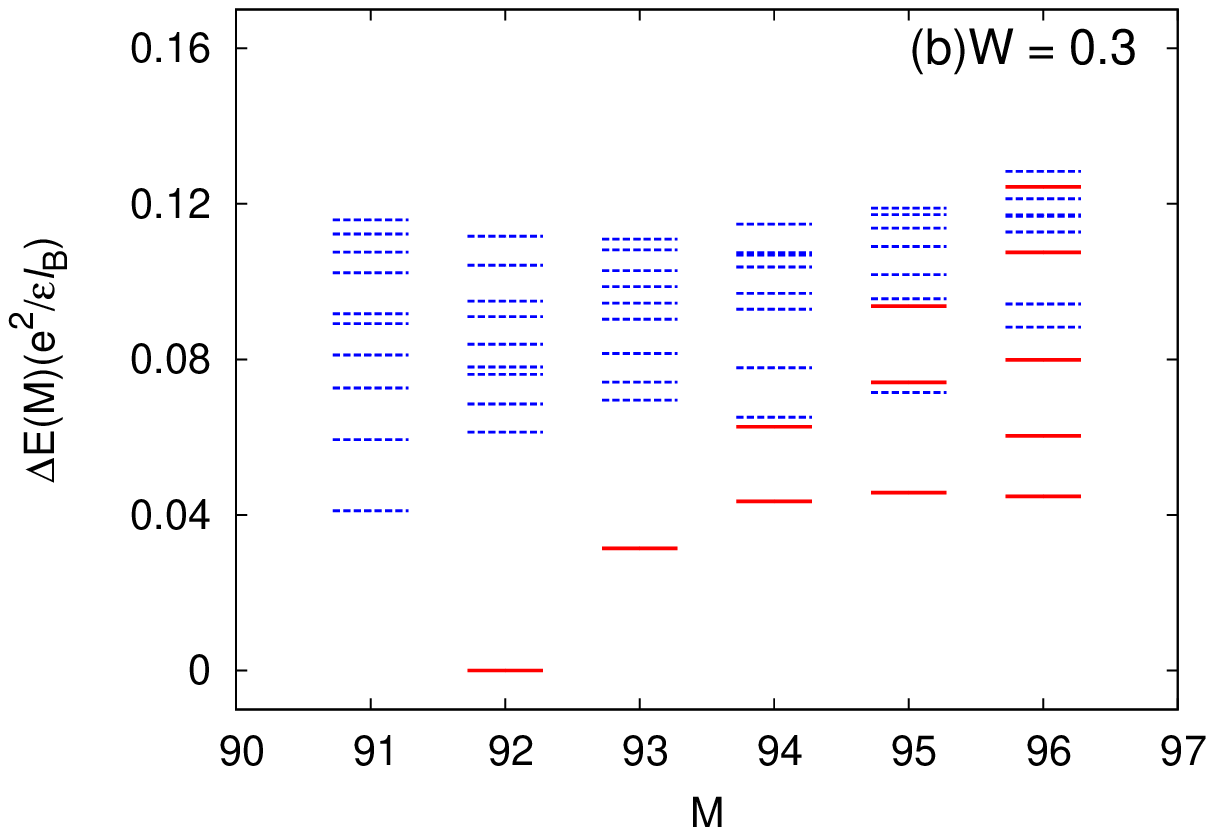}
\caption{\label{edge1} (color online).  (a) The
   low-lying energy states (edge states are marked by solid red bars)
   for 8 electrons in 26 orbitals with Coulomb interaction.
   (b) The energy spectrum after exciting a quasihole
   at the origin by a short range potential $W_0c_0^{\dagger}c_0$
   with $W_0 = 0.3$.}
\end{figure}

In Figure~\ref{edge1}, we compare the low-energy excitations
of the system with and without the quasihole excitation. We identify
the edge excitations, labeled by solid red bars, following the
approach developed by one of the authors and his
collaborators.~\cite{wan03} We observe that in energy relative to
the ground state, the edge spectrum looks almost identical with and
without the quasihole, implying the Abelian nature of the quasihole.
This contrast to the case of a Moore-Read state, where the presence
of a $+e/4$ quasihole changes the fermionic edge excitations. The
number of the edge states (including the ground state) are 1, 1, 2,
3, and 5 for $\Delta M = 0$-4, as expected by the chiral boson edge
theory.~\cite{wen:405,wen:IJMPB}

The short-range potential is useful to generate a single $+e/3$
quasihole if the edge confinement is not too strong. However, since
it only affect the local potential at a single orbital, a second
quasihole cannot be induced, since one cannot deplete more than
(on average) 1/3 charge in a single orbital in the Laughlin case.
For the same reason, the short-range potential does not support
a single quasiparticle (charged $-e/3$), as one can see from
Fig.~\ref{v1profilenum} a quasiparticle occupies several orbitals,
unlike a quasihole. Therefore, we proceed to study local potentials
with a longer range.

\subsection{Gaussian-shaped potential}

We now considered the Gaussian-shaped potential $H_W = W_g \sum_m \exp
(-m^2 / 2 s^2)c_m^{\dagger}c_m$, or the potential has a value
$W_g \exp(-m^2/2s^2)$  on the $m$'th orbital. The width
of the potential is $s$, while the strength of the potential
$W_g$. In the limit of $s \rightarrow 0$, the Gaussian potential
evolves into the short-range $\delta$-potential
discussed in the previous subsection.

\begin{figure}
\includegraphics[width = 8cm]{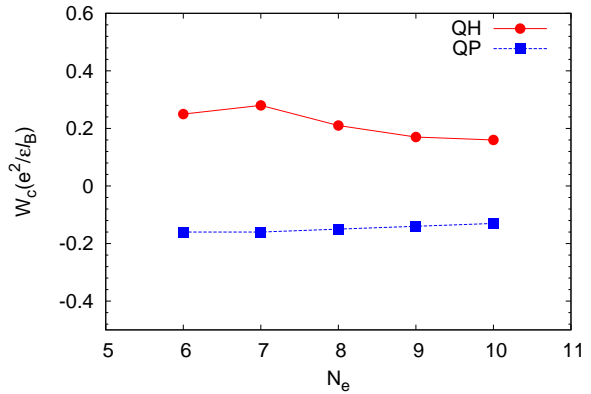}
\caption{\label{wcl1} (color online). The threshold strength
  $W_c$ for the excitation of a quasihole (red dots) or a quasiparticle
  (green squares) in $\nu = 1/3$ systems with 6-10 electrons with
  Coulomb interaction, using
  a Gaussian potential $H_W = W_g \sum_m \exp
(-m^2 / 2 s^2)c_m^{\dagger}c_m$ with width $s = 2$.}
\end{figure}

For fixed $s = 2$ and $d = 0.5 l_B$, we vary $W_g$ to study the
change of the total angular momentum of the global ground state. For
example, in a system of $N = 10$ electrons in 30 orbitals, we find
the total angular momentum $M_{tot}$ jumps from $M_L = 3N(N-1)/2 =
135$ to $M_{1qh} = 3N(N-1)/2 + N = 145$ at $W_g = 0.16 \pm 0.01$,
indicating the presence of one quasihole at the origin. For an
attractive potential, $M_{tot}$ drops from $M_L =135$ to $M_{1qh} =
3N(N-1)/2 - N = 125$ at $W_g = -0.13 \pm 0.01$, indicating the
emergence of a quasiparticle at the origin. Figure~\ref{wcl1}
shows the value of $W_g$ at which $M_{tot}$ changes for systems with
6-10 electrons. We find that the threshold values for the generation
of one quasihole or one quasiparticle approaches a constant value of
roughly $\pm 0.15$ as the system size increases. It is not
surprising that the threshold value is of the same energy scale as 
the bulk energy gap, but it also depends on the detail of the potential.

\begin{figure}
\includegraphics[width = 6.5cm]{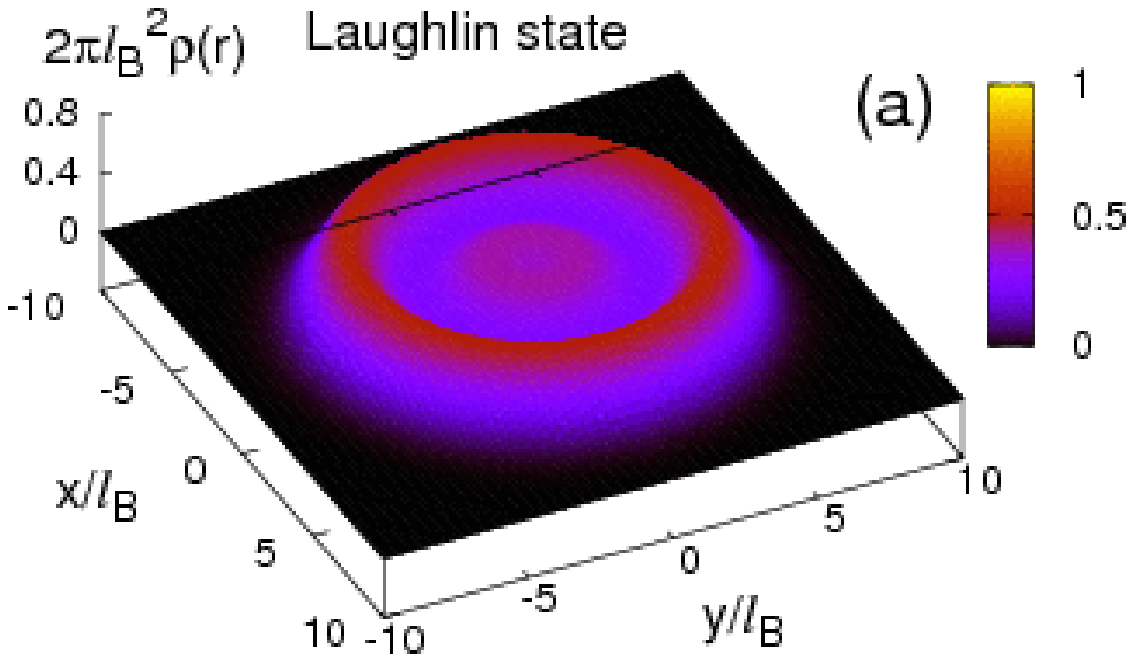}
\includegraphics[width = 6.5cm]{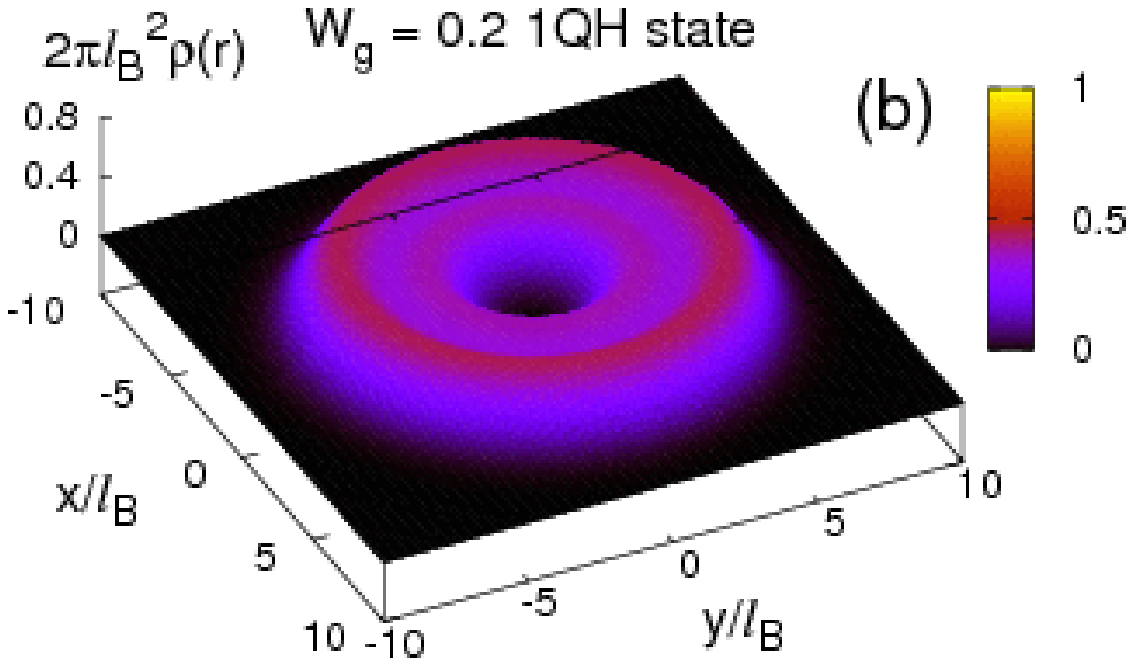}
\includegraphics[width = 6.5cm]{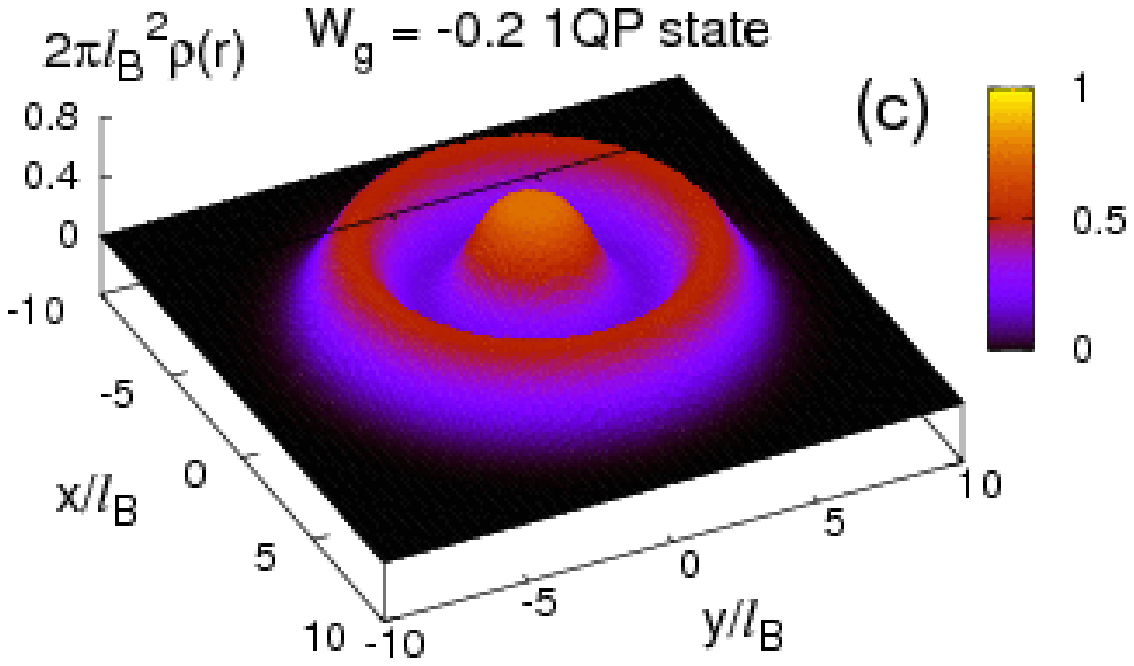}
\caption{\label{gaussiandensity} (color online).
  Electron density profiles for (a) the Laughlin state for $W_g = 0$, (b)
  one-quasihole state for $W_g = 0.2$, and (c) one-quasiparticle state
  for $W_g = -0.2$. We consider a system of 10 electrons in 30 orbitals
  with Coulomb interaction. A Gaussian potential
  $H_W = W_g \sum_m \exp (-m^2 / 2 s^2)c_m^{\dagger}c_m$
  with $s = 2$ is applied.}
\end{figure}

Figure~\ref{gaussiandensity} shows the electron density profiles for
(a) the Laughlin state, (b) a one-quasihole state for $W_g = 0.2$, and
(c) a one-quasiparticle state for $W_g = -0.2$. The density
accumulation or depletion at the origin indicates the presence of a
quasiparticle or quasihole. Compared to the case of the hard-core
potential, the quasihole is slightly larger, while the quasiparticle
is smaller with a well-defined peak at the origin. Therefore, in the
more realistic case with Coulomb interaction (and not too narrow a
tip), the quasihole state and the quasiparticle state have roughly the
same perturbation to the Laughlin ground state except for the opposite
signs, suggesting a quasiparticle-quasihole symmetry.

\begin{figure}
\includegraphics[width=8cm]{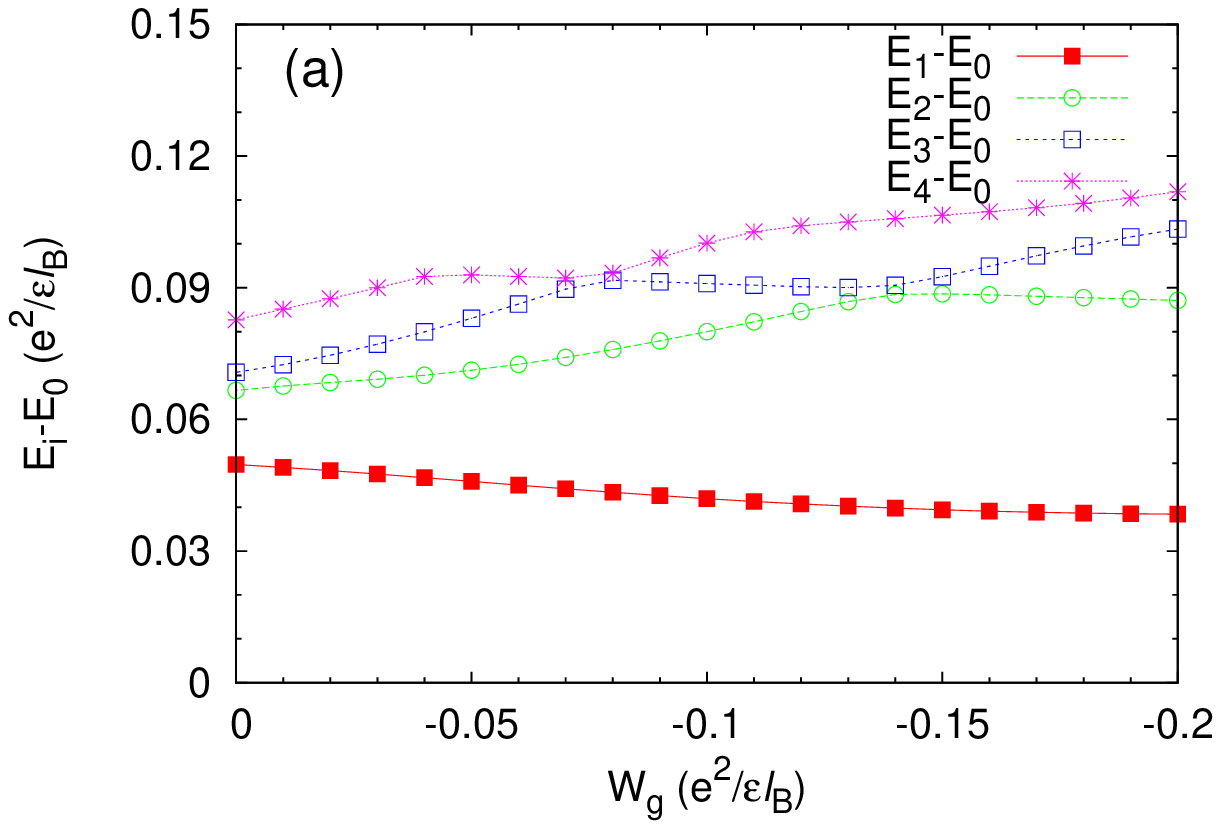}
\includegraphics[width=8cm]{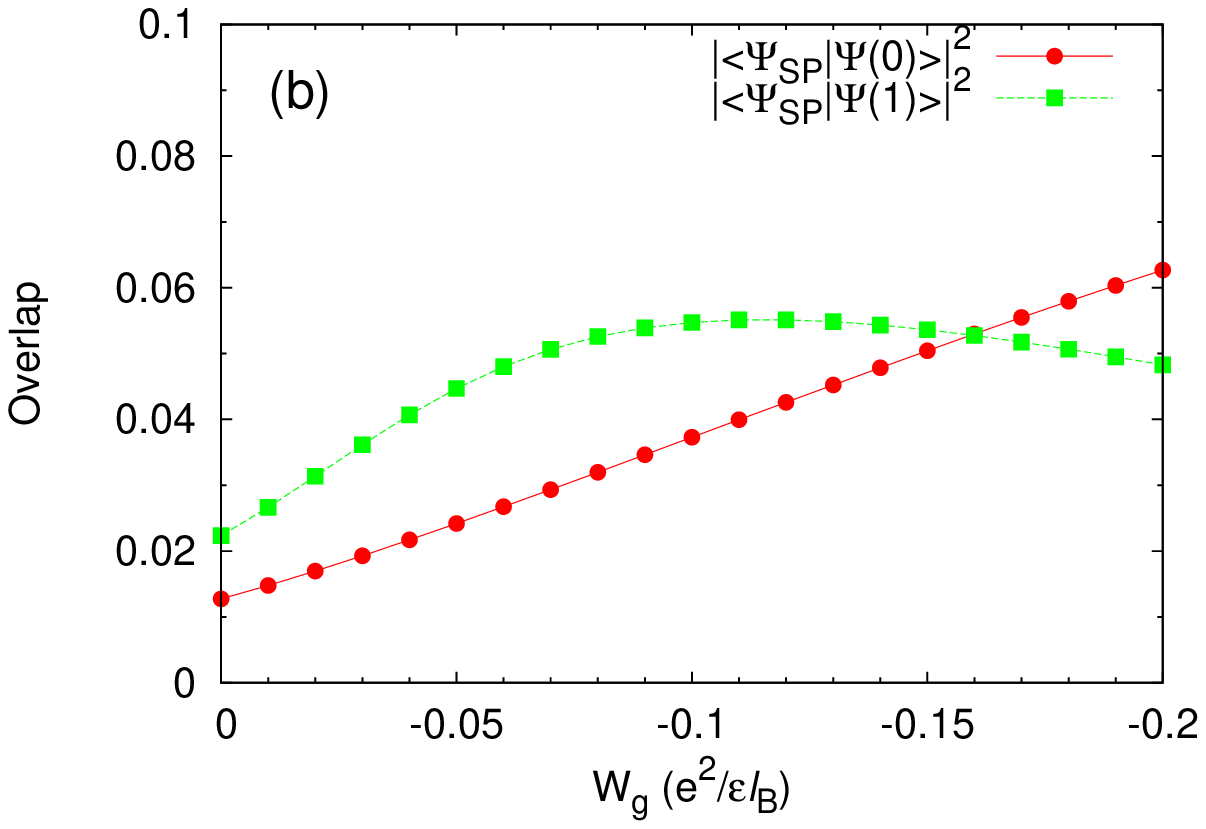}
\caption{\label{evolvem125} (color online).
  Evolution of (a) excitation energies of the lowest four excited states
  and (b) overlaps of the lowest two states with the stripe phase
  $|\Psi_{SP} \rangle = |110000000000111111110000000000 \rangle$.
  The system is of $N = 10$ electrons in 30 orbitals with Coulomb
  interaction ($d = 0.5 l_B$) and a Gaussian potential $W_g \sum_m \exp
  (-m^2 / 2 s^2)c_m^{\dagger}c_m$ with $s = 2$. }
\end{figure}

Since the quasiparticle state we obtain for the Gaussian potential in
systems with Coulomb interaction cannot be easily compared with the
variational wave function [Eq.~(\ref{eqn:quasiparticle})], we want to
make sure it is not a stripe phase, which arises commonly in systems
with Coulomb interaction. In Fig.~\ref{gaussiandensity}(c), we find
the electron density of the quasiparticle state has a large value
around the origin and edge. A stripe phase of $N = 10$ electrons in 30
orbitals with a somewhat similar density distribution and the same
total angular momentum can be represented by a binary string
$|\Psi_{SP} \rangle =|110000000000111111110000000000 \rangle$ (a
Slator determinant), in which each digit specifies the corresponding
single-electron orbitals (from 0 to 29) being occupied (1) or not
(0). Therefore, we wish to answer the question how close the ground
state with $M_{tot} = 125$ is to the stripe phase. For this, we plot
the lowest four excitation energies (energy difference between the
lowest four excited states and the ground state in the subspace of
M=125) as a function of $W_g$ in Fig.~\ref{evolvem125}(a). Obviously,
there is no crossing/anti-crossing between the ground state (which we
identified as the quasiparticle state) and the first excited states as
$|W_g|$ increases. This is very different from the behavior of the
next three excited states, which can get very close in energy. We
further calculated the overlaps between the lowest two energy states
and the stripe state as a function of $W_g$ in
Fig.~\ref{evolvem125}(b). While the overlap is increasing for the
ground state, it is only about 5\% for $W_g \sim W_g^c = -0.13$ when
the ground state in the $M = 125$ subspace becomes the global ground
state. We therefore conclude that the ground state is unlikely the
stripe state.

\begin{figure}
\includegraphics[width=8cm]{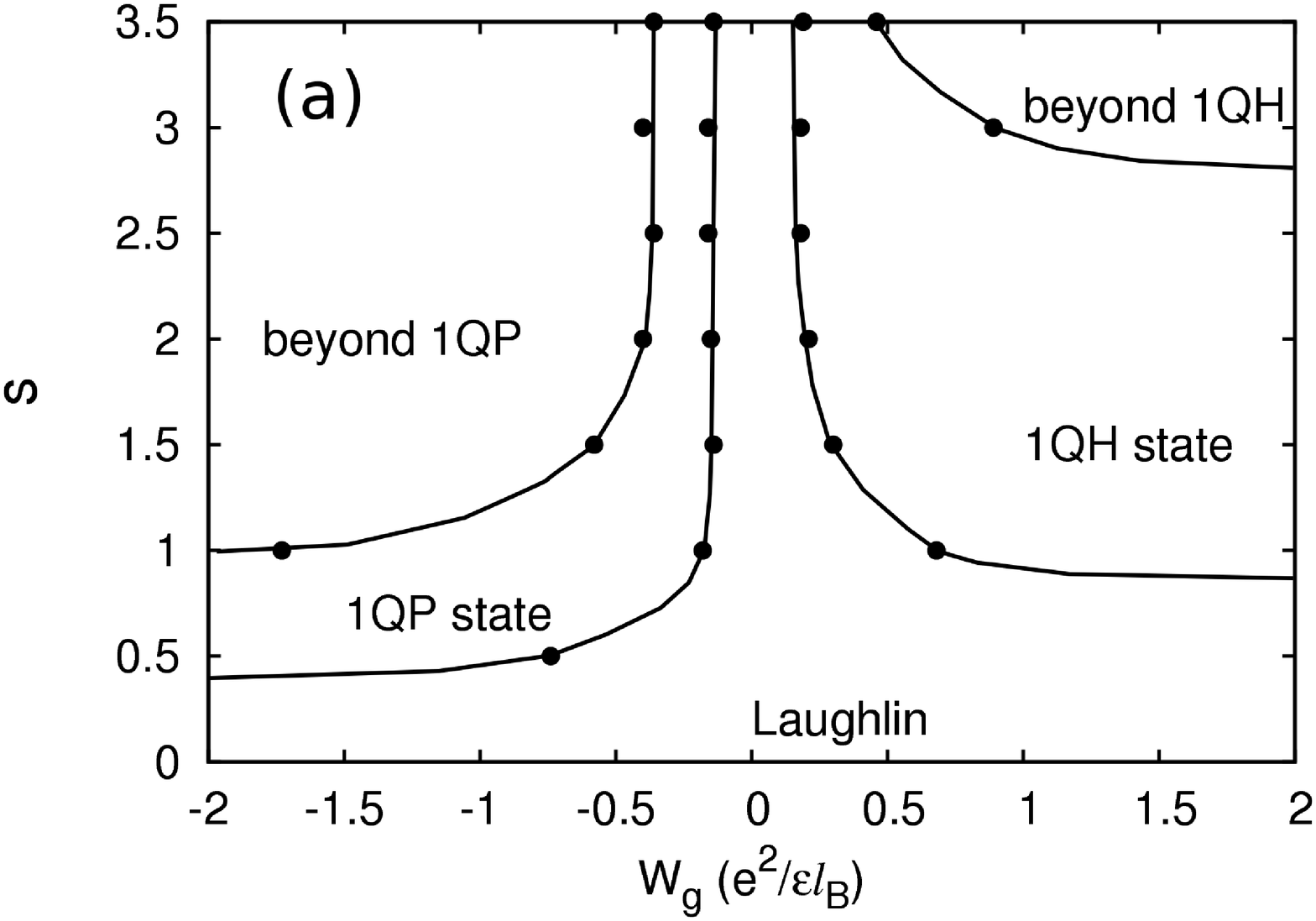}
\includegraphics[width=8cm]{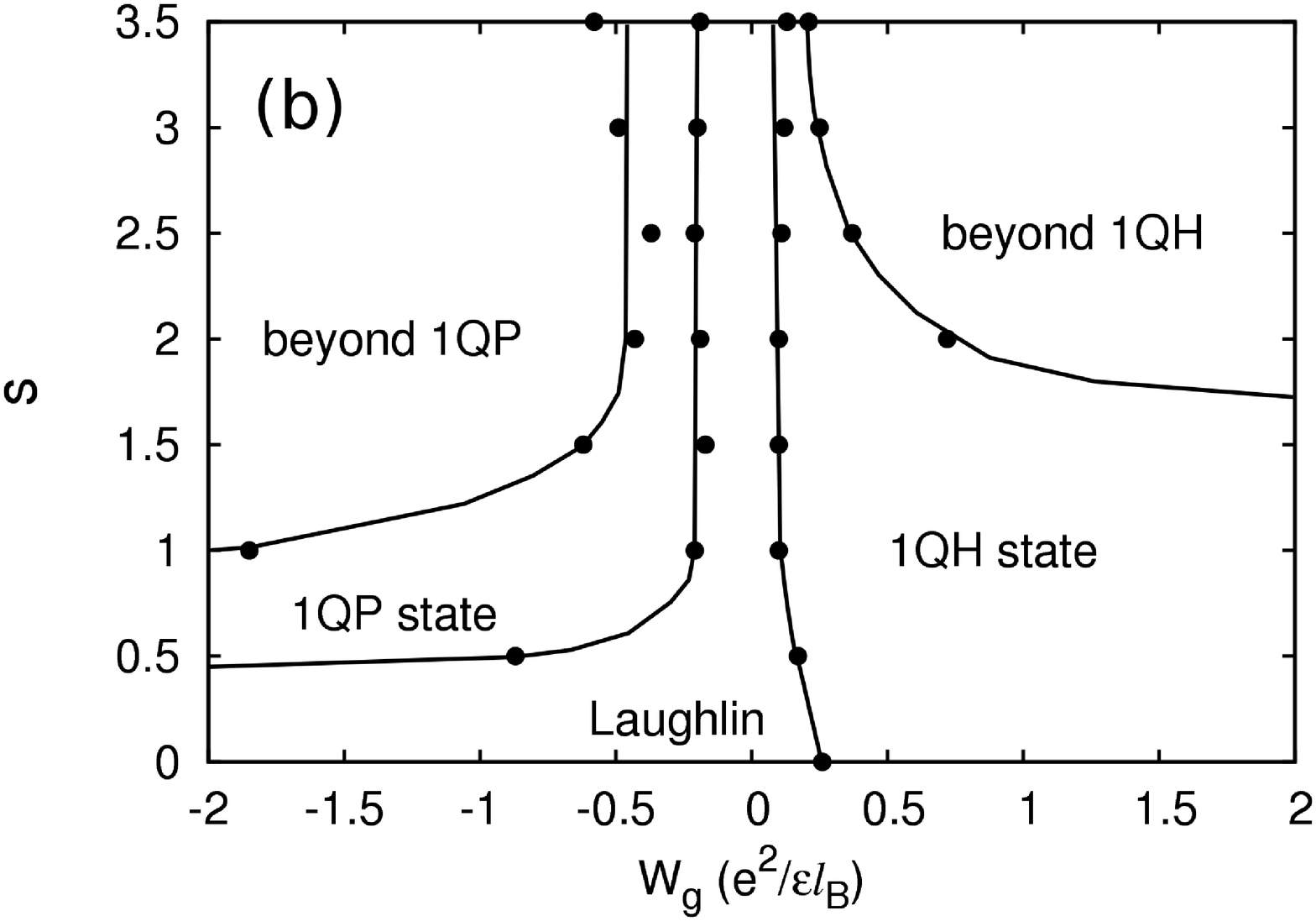}
\caption{\label{phase}
Ground state diagram for the system of 8 electrons in 24 orbitals
with Coulomb interaction, in the presence of the Gaussian tip potential
with strength $W_g$ and width $s$.
The confining charge is located at a distance of (a) $d = 0.5 l_B$
and (b) $d = 1 l_B$.}
\end{figure}

We extend our calculation to a grid on the area defined by $-2 \leq
W_g \leq 2$ and $0 < s < 3.5$ for the Gaussian potential $W_g
\sum_m \exp (-m^2 / 2 s^2)c_m^{\dagger}c_m$.  We choose the strength of
edge confinement to be $d = 0.5 l_B$ (stronger confinement) and $d = 1
l_B$ (weaker confinement).  The results for 8 electrons are plotted in
Fig.~\ref{phase}.  Generically we can devided the parameter
space into five regions: the Laughlin state ($M_{tot} = 3N(N-1)/2$),
the one-quasihole state ($M_{tot} = 3N(N-1)/2 + N$), the
one-quasiparticle state ($M_{tot} = 3N(N-1)/2 - N$), beyond
one-quasihole state (ground states with $M_{tot} > 3N(N-1)/2 + N$),
and beyond one-quasiparticle state (ground states with $M_{tot} <
3N(N-1)/2 - N$).  We have also done the calculation for $N = 10$ with
similar results, but on a coarser grid.  In particular, we do observe
the ground state with $M_{tot} = 3N(N-1)/2 + 2N = 155$ for $N = 10$,
consistent with the angular momentum for a two-quasihole state.  We do
not find any global ground state with $M_{tot} = 3N(N-1)/2 - 2N = 115$
(consistent with that of a two-quasiparticle state), but at 117.  One
might tempt to speculate this as one of the two quasiparticles move
away from the origin.  Nevertheless, in such small systems, it is
unnecessary and most likely unreliable to emphasize multiple
quasiparticle and quasihole excitations, so we simply mark the regions
with $M_{tot} > 3N(N-1)/2 + N$ and $\Delta M_{tot} < 3N(N-1)/2 - N$ by
``beyond 1QH'' and ``beyond 1QP'', respectively, and do not proceed
further.

The main difference between $d = 0.5 l_B$ and $d = 1.0 l_B$ occurs
along the boundaries of quasiholes, not along the quasiparticle
boundaries.  This, we believe, is due to the fact that for a fixed
number of electrons the quasihole states (not the quasiholes
themselves) have larger size than the quasiparticle states, thus more
susceptible to the edge confinement. The difference is more evident at
smaller $s$ (sharper tips). In particular, a $\delta$-tip can excite a
quasihole in the case of $d = 1.0 l_B$, but not in the case of $d =
0.5 l_B$ for not too large $W_g$. This, as illustrated in
Fig.~\ref{phase}, suggests that a finite width $s \approx 2 l_B$ may
be more robust for the excitation of quasiholes and quasiparticles.

\subsection{Exponential-shaped potential}

In this section we discuss the exponential-shaped potential
$H_W = W_e \sum_m \exp (-m /\xi)c_m^{\dagger}c_m$.
In real space, this corresponds to a Gaussian potential
$V(z) = W_g^r e^{-|z|^2/2\sigma^2}$, which may not be
too difficult to prepare in experiments.
Explicitly, by projecting the real-space potential into the
lowest Landau level, we obtain the matrix elements
\begin{eqnarray}\nonumber
  \langle \phi_m|V|\phi_m \rangle &=& \frac{W_g^r}{2\pi 2^m m!}
  \int_0^\infty   e^{-\frac{l_B^2+\sigma^2}{2\sigma^2}{|z|^2 \over l_B^2}}
\frac{|z|^{2m}d^2z }{l_B^{2m+2}}
\\\nonumber
  &=& W_g^r \left (\frac{\sigma^2}{l_B^2+\sigma^2} \right )^{m+1} \\
 &=& W_e e^{-m /\xi},
\end{eqnarray}
where the decay length is $\xi = 1 / \ln (1+l_B^2/\sigma^2)$ and
the effective strength $W_e = W_g^r \sigma^2/(\sigma^2 + l_B^2)$.
Again, $|\phi_m \rangle$ is
the lowest landau Level wave function with angular momentum $m$.

\begin{figure}
\includegraphics[width=6.5cm]{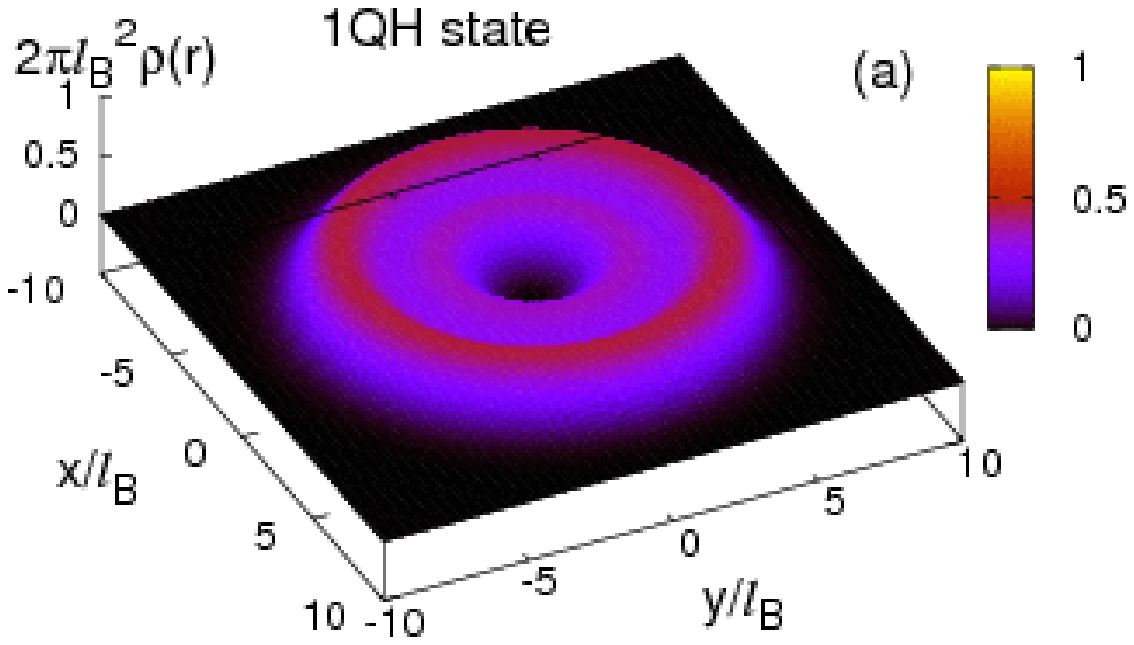}
\includegraphics[width=6.5cm]{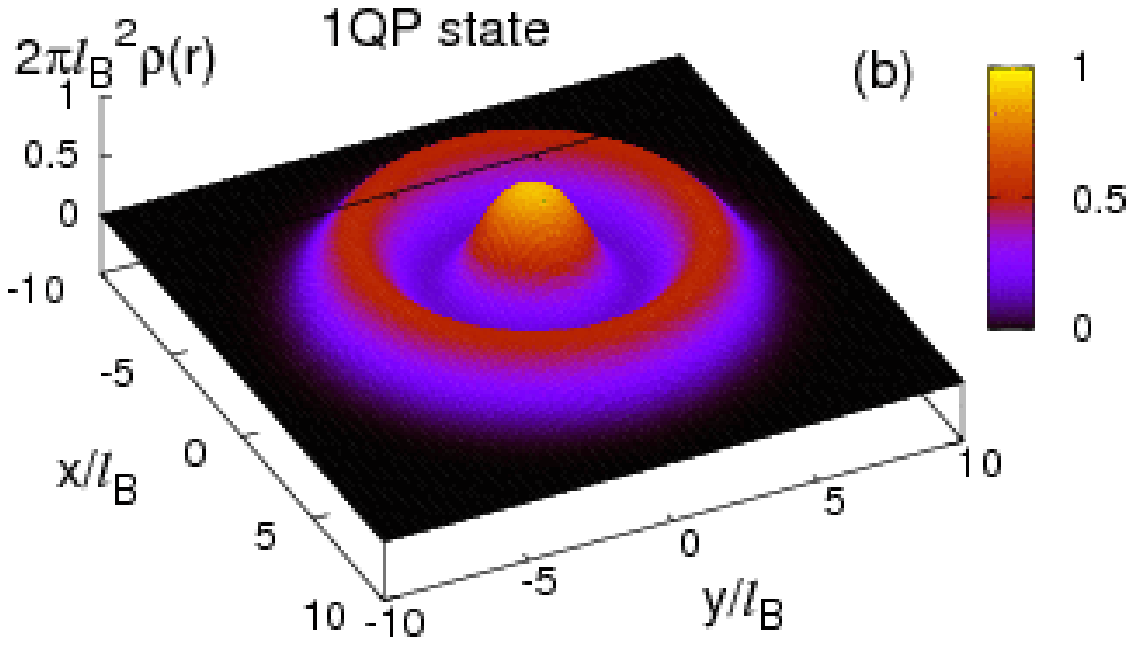}
\caption{\label{expdensity} (color online).  The electron density
  profiles for (a) the one-quasiparticle state ($W_e$=-0.2) and (b)
  the one-quasihole state ($W_c$=0.28) for the exponential potential
  $H_W = W_e \sum_m \exp (-m /\xi)c_m^{\dagger}c_m$ with $\xi = 1/\ln
  2$ (or $\sigma = l_B$ in real space).  The system has 10 electrons
  in 30 orbitals with Coulomb interaction.  The neutralizing confining
  charge is located at a distance of $d = 0.5 l_B$ above the electron
  layer.  }
\end{figure}

After applying the exponential potential $H_W = W_e \sum_m \exp (-m
/ \xi)c_m^{\dagger}c_m$ with $\xi = 1/\ln 2$ (or $\sigma = l_B$ in
real space), we are also able to trap a single quasihole or a
quasiparticle. Again, we consider a system of 10 electrons in 30
orbitals, with neutralizing confining charge located at a distance
of $d = 0.5 l_B$ above the electron layer. The electron density
profiles for the Laughlin state, the one-quasihole state, and the
one-quasiparticle state are plotted in Fig.~\ref{expdensity}). The
density profiles look very similar to those for the Gaussian
potential discussed in the previous subsection (Fig.~\ref{gaussiandensity}). Like the Gaussian case, the quasiparticle state and
the quasihole state have roughly the same density perturbation (but
with opposite signs) to the Laughlin ground state.

\section{conclusion and discussion}

To summarize, we study the trapping of quasiholes and quasiparticles
by a local potential (e.g. induced by an AFM tip) in a microscopic
model of fractional quantum Hall liquids with short-range hard-core
interaction or long-range Coulomb interaction with an edge confining
potential due to neutralizing charge.  We find, in particular, at the
Laughlin filling faction $\nu = 1/3$, both quasihole and quasiparticle
states can be energetically favorable for the ground state of the
Coulomb system for tip potentials of various shape and strengths. The
presence of the Abelian quasihole has no effect on the edge spectrum of
the quantum liquid, unlike in the non-Abelian case when fermionic
excitations are present.

Although quasiholes and quasiparticles can emerge generically in the
system, its stability depends on the strength of the confining
potential, the strength and the range of the tip potential.
Experimentally the quantum Hall plateau at $\nu = 1/3$ was found in a
high magnetic field ($\sim 15$ T).~\cite{PhysRevLett.48.1559} In this
case the magnetic length $l_B \approx 70$ \AA. Based on our
microscopic calculation, we estimate a optimal range of the tip
potential to be 140 \AA. The size falls in the right range of AFM tip
size under current technology. The Laughlin state in the context of
topological quantum computing is of less interest due to its Abelian
nature, although it can be used for topological quantum
memory. Nevertheless, it is much easier to model in numerical studies
than the non-Abelian Moore-Read state,~\cite{wan07} and the even more
complicated Read-Rezayi (parafermion) states.~\cite{PhysRevB.59.8084}
We expect the results found here can be of help for the excitation and
trapping of quasiholes or quasiparticles in the Moore-Read case in future
experiments. In the Moore-Read case at filling fraction of 5/2, a
smaller magnetic field $\sim 5$ T is usually applied. Thus with a
longer magnetic length we can have even wider tips, which should not
be a technical challenge.

With the well-known difficulties of the exact diagonalization method
in highly entangled systems such as the fractional quantum Hall
liquids, the search for the ground states with a few parameters is a
time-consuming job.  The Moore-Read case is even more complicated,
since the even-denominator state has a smaller excitation gap and is
competing with stripe phases.~\cite{wan07} One might wish to develop
more efficient numerical methods to approach the ground state
properties.  One development in recent years is the application of
density-matrix renormalization group (DMRG)
method~\cite{PhysRevLett.86.5755,
  J.Phys.A.36.381,J.Phys.A.17.7335,Feiguin07} to the fractional
quantum Hall systems. We implememt the method in the disk geometry
with results in excellent agreement with exact diagonalization in
small systems.~\cite{future} However, we find the time to reach
convergence (especially near the origin) in larger systems is
impractically long for the extensive search for the ground states
discussed in the current paper.

\begin{acknowledgments}

  We thank R.~R. Du for the illuminating discussion on the latest
  experimental aspects of high-quality two-dimensional electron gases.
  X.W. also benefits a lot from the collaboration with Ed Rezayi and
  Kun Yang on related projects.  We acknowledge the support from the
  National Natural Science Foundation of China through Grant
  No. 10504028. This research was supported in part by the PCSIRT
  (Project No. IRT0754) and by the PKIP of CAS.  Z.X.H. thanks the
  CCAST for hospitality during a joint workshop with the KITPC on
  ``Topological Quantum Computing'' in Beijing.

\end{acknowledgments}

\end{document}